\documentclass[lettersize,journal]{IEEEtran}
\usepackage{amsmath,amsfonts}
\usepackage{algorithmic}
\usepackage{algorithm}
\usepackage{array}
\usepackage{subcaption}
\usepackage{graphicx}
\usepackage{textcomp}
\usepackage{stfloats}
\usepackage{url}
\usepackage{verbatim}
\usepackage{cite}
\usepackage[normalem]{ulem}
\usepackage{svg}
\usepackage{tcolorbox}
 \usepackage{balance}
\newcommand{\explains}[2]{\begin{footnotesize}{\textsf{\textbf{#1.}~#2}}\end{footnotesize}}

\begin{document}

\title{Physiological Data: Challenges for Privacy and Ethics}

\author{Keith M. Davis III$^1$ and Tuukka Ruotsalo$^{2,3}$
       \\
       first.last@helsinki.fi, first.last@lut.fi \\ 
       $^1$University of Helsinki, Finland\\
       $^2$University of Copenhagen, Denmark\\
       $^3$LUT University, Finland\\
\textbf{© 2024 IEEE.} Personal use of this material is permitted. Permission from IEEE must be obtained for all other uses, in any current or future media, including reprinting/republishing this material for advertising or promotional purposes, creating new collective works, for resale or redistribution to servers or lists, or reuse of any copyrighted component of this work in other works.
\thanks{The content of this preprint has been accepted for publication in IEEE Computer, DOI: 10.1109/MC.2024.3404994.}

}

\markboth{Davis and Ruotsalo. Physiological Data: Challenges for Privacy and Ethics}%
{Shell \MakeLowercase{\textit{et al.}}: A Sample Article Using IEEEtran.cls for IEEE Journals}


\maketitle
\begin{figure*}
    \centering  \includegraphics[width=0.80\textwidth]{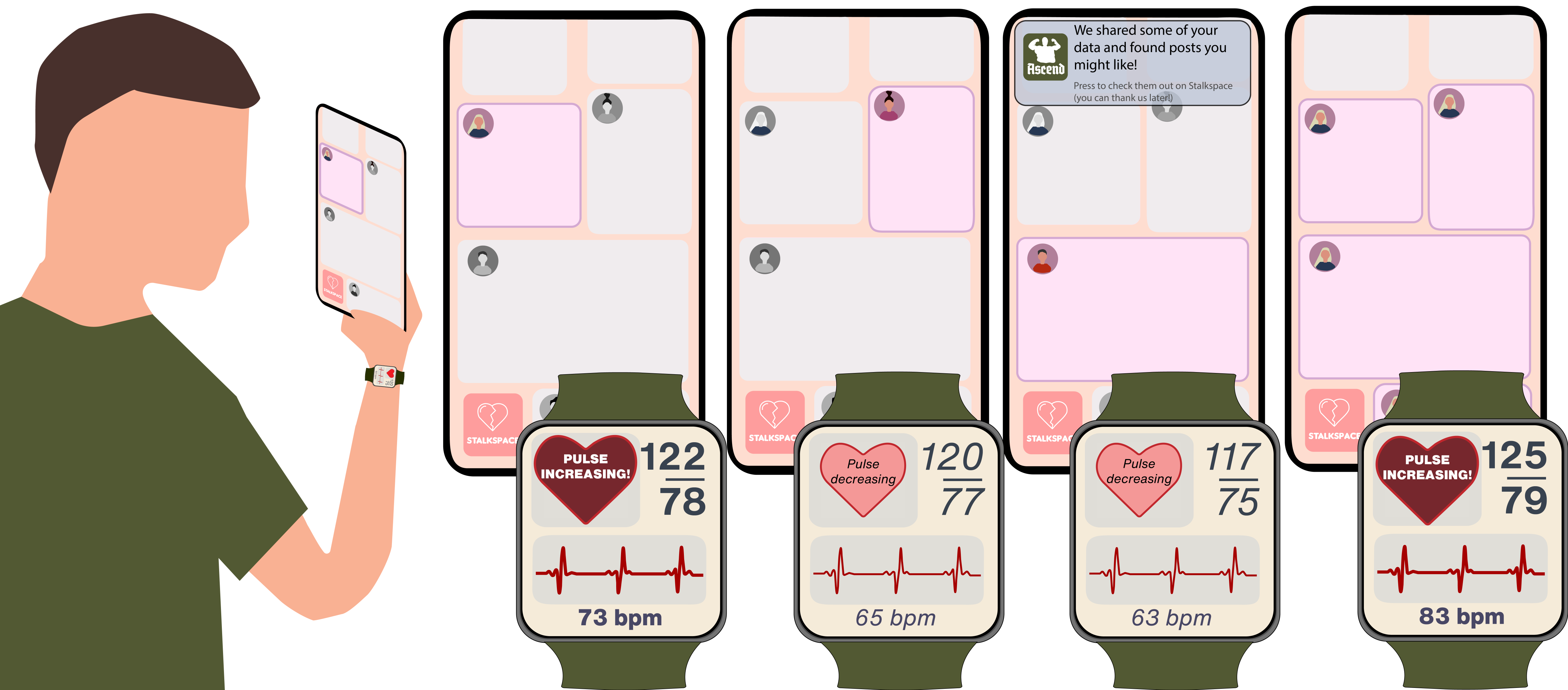}
  \caption{A person who intends to use physiological devices may be unaware that the signals collected from these devices can be paired with data collected from their smartphone. In the scenario depicted above, a user browsing social media while wearing a smartwatch discovers the watch is sharing his physiological data with the social media service he is using. His heart rate can then be used to determine which content he finds most interesting or appealing. While some users may appreciate such a service, others are likely to find it invasive.} 
  \label{fig:teaser}
\end{figure*}

\begin{abstract}

Wearable devices that measure and record physiological signals are now becoming widely available to the general public with ever-increasing affordability and signal quality. The data from these devices introduce serious ethical challenges that remain largely unaddressed. Users do not always understand how these data can be leveraged to reveal private information about them and developers of these devices may not fully grasp how physiological data collected today could be used in the future for completely different purposes. We discuss the potential for wearable devices, initially designed to help users improve their well-being or enhance the experience of some digital application, to be appropriated in ways that extend far beyond their original intended purpose. We identify how the currently available technology can be misused, discuss how pairing physiological data with nonphysiological data can radically expand the predictive capacity of physiological wearables, and explore the implications of these expanded capacities for a variety of stakeholders.

\end{abstract} 


\IEEEPARstart{A}{dopting} novel technologies before understanding the extent of their consequences is a theme that has persisted throughout much of human history. The early internet, limited by the computational technologies available at the time, was devoid of much of the widespread automation experienced today. It was taken for granted that the user experience, for better or worse, was created by other humans. 

Today, the Internet is hardly recognizable from its humble origins, having been dramatically transformed by ubiquitous Internet-enabled smartphones, social media, inexpensive computational power, and large advances in machine learning and artificial intelligence. The collective effects of these advances have led to significant changes in how organizations, from small businesses to governments, operate. The wellspring of ubiquitous user data, especially user data that can be connected with social media profiles, has significantly expanded the ability of organizations with access to such data to understand human behavior. However, the present developments of ubiquitous data have mostly been under the intentional control of the users that produce them. If a user wishes to keep their personal opinions hidden, they can do so by not publicly engaging with content.

Like the Internet, physiological wearables have also changed dramatically in the past two decades. Advances in computational power, battery lifespan, chip size, and wireless connectivity have facilitated the manufacturing of lightweight devices capable of measuring multiple signal sources simultaneously for extended periods without having to be removed for charging while transmitting these data sources wirelessly to smartphones and other Internet-connected devices.

While portable sensor technology facilitating highly accurate real-time monitoring of physiological signals presents new and exciting opportunities for individuals to better understand themselves, it also creates novel ways to violate individual rights to privacy and autonomy. The capacity to model human preferences and behaviors at unprecedented scales has already radically altered how content, products, and even worldviews are marketed to the general public. These capacities will continue to expand as physiological computing becomes more pervasive and may present serious and permanent threats to the public at large if they remain unaddressed.


\section{Promises and Perils}

Large internet service companies and a seemingly endless cornucopia of data peddlers and merchants have already demonstrated that low-fidelity data collected from millions of users can be used to infer complex behavioral patterns from individual users. Although the higher-order emergent consequences of these new digital marketing and interaction paradigms created by such companies have yet to be fully realized, their effects on society are becoming more apparent with every passing year. A newer paradigm with similar properties is emerging: ubiquitous physiological computing, paired with ubiquitous behavioral data.

While increased access to one's physiological information could confer significant benefits to an individual, the scope of the possible negative consequences of ubiquitous high-throughput, low-fidelity physiological data rapidly expands upon closer scrutiny. Previous work has engaged with weighing the risks and benefits of physiological data from the perspective of health and medical research, \cite{nebeker2019building, torous2017ethical}, but the subject of pairing physiological and non-physiological in everyday applications has not been adequately explored.

The potential issues created by these devices and the data they produce can exist within many different degrees of separation tied directly to their usage, ranging from those stemming directly from the devices themselves to emergent effects that do not become fully apparent until long after the devices become a part of daily life. First-order problems, where negative effects can be more directly linked to the devices and applications themselves (such as excessive and compulsive checking of health metrics provided by a device), may be relatively easy to find direct solutions for. On the other hand, more complex issues produced by the interaction of data produced by physiological wearables and associated data from other digital applications and services are more challenging to diagnose and solve.

Combining physiological data collected from various devices worn by users with their existing digital information, as depicted in Figure \ref{fig:teaser}, could grant any organization enormous advantages in understanding a user's behaviors, affective states, and cognitive states. These advantages would not come simply from having more data, but from having access to data that may predict patterns and behaviors that non-physiological data cannot.

For example, heart rate variability (HRV), measured in isolation, currently serves little more than as a metric to gauge cardiovascular health. However, pairing it with browsing behavior opens a wellspring of potential future abuses of user privacy, given that these physiological signals can be used to estimate the emotional state of the individual. Under such conditions, an individual's private opinions and feelings towards content could be assessed without their knowledge or consent. It also reveals a fundamental problem with physiological data: collected signals previously considered innocuous can later pose problems for user privacy if they contain additional latent information.


Concerns related to ubiquitous computing for privacy and privacy models designed to contend with these concerns are not new and have been discussed at length in previous work \cite{langheinrich2001privacy, hong2004privacy, nissenbaum2004privacy}. However, discussion is fairly limited regarding how to deal with the novel paradigm introduced by physiological wearables, where instead of using conscious actions performed by a user (e.g. clicks, likes, shares, and comments) to build a profile of their interests and preferences, a user profile can instead be built through passive monitoring of a user's physiological signals.

In Nissenbaum's \textit{Privacy as Contextual Integrity}\cite{nissenbaum2004privacy}, the primary framework offered for conceptualizing privacy first requires a reasonable understanding of how certain data can be used so informed decisions can be made regarding who should and should not have access to them. But physiological data have not been well-studied in big data or multimodal contexts, thus their full predictive capacities and potential use cases are currently unknown. Additionally, common across the privacy literature are suggestions to contend with the privacy risks of emerging technologies by designing them with integrated privacy-preserving protections and restrictions. Unfortunately, the always-on nature of physiological wearables and the data they collect makes it difficult to create truly "privacy-centric" designs.

The advent of ubiquitous physiological data presents unprecedented challenges to ethics and privacy. Careful consideration and rigorous investigation are necessary to better understand how these new data streams might be used in ways we cannot yet anticipate, and how existing privacy frameworks can be adapted or augmented to address these novel challenges.

\begin{figure*}
\begin{tcolorbox}[colback=gray!5!white, colframe=gray!75!black, title=Glossary, fonttitle=\bfseries]
\explains{Data Steward}{A data steward is any party that may collect, handle, distribute, and/or use data collected from users. This is distinct from established terminology common in both legislation and business contexts (such as GDPR's "data controller"). We have selected this term to simplify the discussion and to avoid dependency on definitions specific to legislation.}\\
\explains{Behavioral data}{Behavioral signals are observable interactions with computing systems. They can be \emph{explicit} interactions such as sharing or liking content, but can also be measured \emph{implicitly} as side-information of a primary activity. For example, click-through data to monitor which links users follow and dwell time to measure how long users spend on content.} \\
\explains{Physiological data}{Psychophysiological processes often directly relate to how the human body and brain react to psychological states or external events. Physiological data are the stored measurements of the biological signals produced by these processes.}\\
\explains{Coerced Utilization}{When social, economic, or legal pressures create a strong incentive to adopt and use a technology to obtain social or material advantages and/or to avoid social or economic disadvantages from refusing to adopt and use the technology.}\\
\explains{Primary use}{When physiological data are used in their original context. Carries the lowest risks for abuse and misuse.}\\
\explains{Secondary use}{When physiological data are used for other purposes beyond the original scope, but without supplement from other data.}\\
\explains{Auxiliary use}{When physiological data are combined with other sources of data to be used in contexts well beyond the original intention of the user. Has the greatest risks for abuse and misuse.}\\
\end{tcolorbox} 
\end{figure*}

\subsection{Devices and Signals}

Physiological wearables have rapidly grown in popularity over the past decade. Fitness watches, sleep monitors, and portable brain-monitoring headsets are just some of the equipment available and affordable to the general public. Currently, they are primarily marketed for and used within the context of health and wellness, such as for exercise and sleep tracking. While early wearable fitness devices only provided basic measurements such as heart rate and estimations of calories burned, more recent devices are capable of performing more advanced tasks, including recording electrocardiograms (ECG), blood pressure, and electrodermal activity (EDA).

Accelerometers and heart rate sensors, beyond simply identifying cardiovascular disorders, can be used to detect psychological conditions and illnesses like depression, bipolar disorder, schizophrenia, post-traumatic stress disorder, and Parkinson's disease \cite{reinertsen2018review}. Neurophysiological signals can be used to diagnose many of these disorders, such as Alzheimer's, schizophrenia, and epilepsy, and can also be used to detect physical illnesses, such as the presence of tumors. There is also some evidence suggesting attention disorders such as attention deficit hyperactivity disorder (ADHD) can be detected using EEG (electroencephalography), although the reliable use of EEG as a diagnostic tool is still debated \cite{arns2013decade}.
 
Wearables provide information that can be used in contexts outside of just detecting or monitoring health conditions. Both heart rate variability (HRV)\cite{ reinertsen2018review} and EEG signals\cite{mikhail2010emotion} can be used independently to detect and classify emotional states, and are fairly accurate even when the data contains a high amount of artifacts and noise. When combining EEG with facial video, such emotional state classification can be done in real-time and with high accuracy \cite{soleymani2014continuous}. 
An obvious application of collecting information related to brain activity outside of predicting seizure episodes or detecting pathologies is to assist in activities that are correlated with changes in cognitive states. Indeed, EEG has been successfully applied to mindfulness and meditation applications \cite{kosunen2016relaworld}, as well in brain-computer interfacing (BCI) contexts for information retrieval, content recommendation, and even to control generative models \cite{spape2021brain}. 

EEG has a long history of being used for guilty knowledge detection \cite{boaz1991detection, merzagora2006wavelet, abootalebi2009new}, as has skin conductance \cite{giesen1980guilty}. While models designed to detect falsification of information using only brain signals are not perfect, perfect performance is not necessary to pose significant privacy concerns. After all, a model capable of accurately classifying a very small fraction of all incoming signals with high confidence can still reveal details about a person that they may wish to otherwise keep hidden.

While it is outside the scope of this article, Lopez et al. \cite{lopez2020beyond} provide a more thorough review of the most likely fields in which commercial EEG devices could be applied. These include medicine, self-regulation and enhancement, smart environments, games and other forms of entertainment, education, and security. These categories could also be generalized to other wearables that monitor other physiological signals. A final category, neuromarketing, has also been proposed, although physio-marketing may be more appropriate, owing to how many physiological signals, not just those recorded from the brain, may be used for purposes of market research.

Physiological monitoring applications have already been introduced by state organizations and private companies in the People's Republic of China with the intention of boosting worker productivity and improving the learning outcomes of students \cite{lopez2020beyond}. Such applications, however, have been met with skepticism and caution, and there is certainly reason to be concerned regarding the early adoption of this technology.

Physiological data can take many forms and can be combined with many other types of data for purposes of analysis and prediction by different stakeholders. Moreover, the data can be used for a wide span of applications, from simple monitoring of an individual's physiological states to predictions of experience and affective stance towards digital and physical environments\cite{10506106}. How physiological data are used and potentially combined with other data ultimately determines their capacity to create serious moral and ethical concerns. 

\begin{figure*}%
\centering
\begin{subfigure}{.90\columnwidth}
\includegraphics[width=\columnwidth]{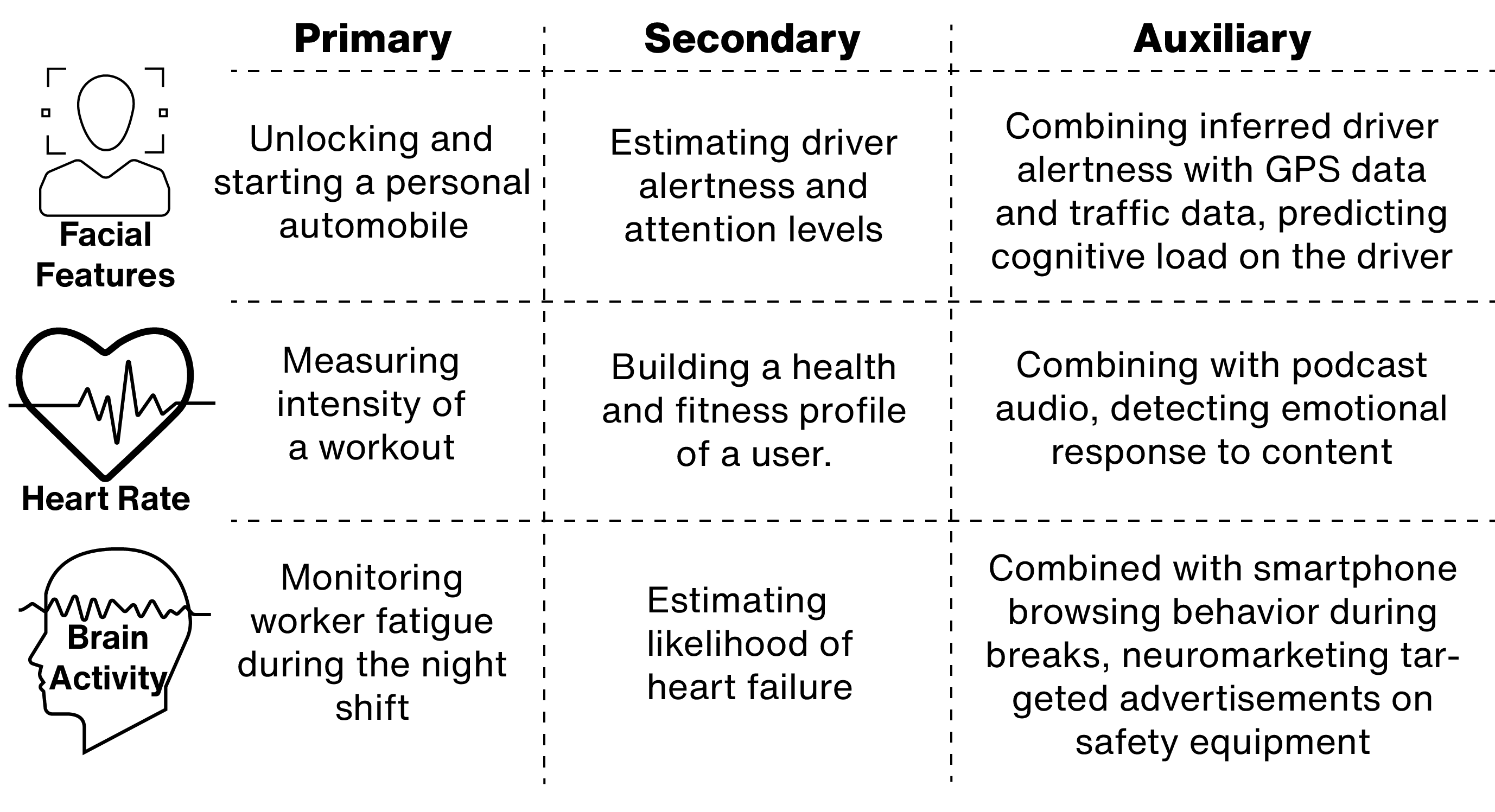}%
\caption{Examples of physiological data usage scenarios.}
\label{fig:sub1}
\end{subfigure}\hfill%
\begin{subfigure}{.98\columnwidth}
\includegraphics[width=\columnwidth]{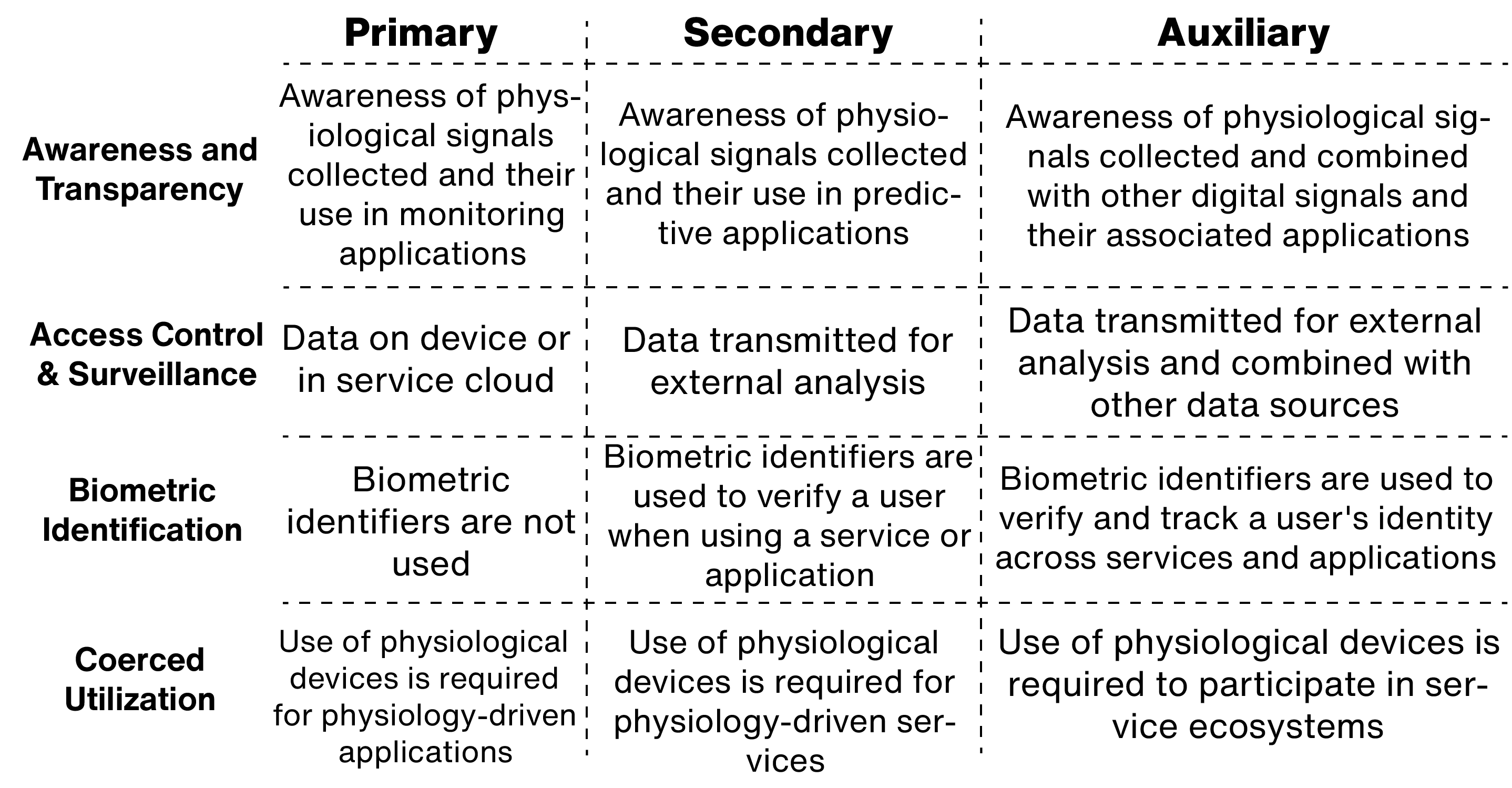}%
\caption{Categories of concern across various data scenarios.}
\label{fig:sub2}
\end{subfigure}%
\caption{Physiological signals come in many forms, and their utility can expand dramatically depending on how they are analyzed and what other data they are combined with.}
\label{fig:hasil}
\end{figure*}

\subsection{Primary Use}

While it may be the expectation that software applications should be free from security flaws and other vulnerabilities, protecting user data is even more important when it involves physiological signals. Previous work has focused on identifying core themes for user privacy concerns, such as in \cite{motti2015users}, where concerns were identified between wrist-worn devices and those worn on the head. Given the variety of characteristics that can be inferred from an individual using a relatively small amount of their physiological data, accidental exposure of this information to third parties has consequences that are far more severe than a similar exposure of most non-physiological data sources.

Considering security, the main sources of vulnerability within primary use scenarios are the app developers and the manufacturers/developers of the devices. Data leaks of raw physiological data, whether facilitated by hackers or incompetent designers, can threaten user privacy. The security flaws discovered in early \textit{smart} devices, such as home appliances, demonstrate the necessity to prioritize security during development. 

Given the widespread adoption of physiological sensors, it is not unreasonable to assume many software applications could benefit from access to the data these sensors provide. For any given individual, this could mean dozens or even hundreds of software applications for which they have granted access to their physiological data, thus greatly increasing the likelihood of their physiological data being eventually exposed.

Manufacturers of physiological wearables face challenges that are distinct from those encountered by manufacturers of other types of electronic peripherals. Concerning development and API support, wearables may fall roughly into two categories: integrated and standalone. Integrated devices, such as the Oura ring\footnote{https://ouraring.com}, operate within a proprietary application ecosystem before any of the physiological data can be accessed through third-party applications. Here, potential exposure of user data can stem from flaws in how the proprietary software applications are designed. For wearables that do not come with their own service applications, but instead provide sensor data and a basic API designed to be used with applications created by third-party developers, vulnerabilities in the API itself and the third-party applications using the data API could be exploited to expose user data. Wearables that allow unrestricted access to raw signal data leave vulnerable an even larger amount of user information that could compromise the user's privacy.

\subsection{Secondary Use}

Even if perfect security of physiological data could be guaranteed, such data contains additional information about a user that they may otherwise wish to keep private. Physiological monitoring, without other sources of information, can be used to construct a comprehensive physiological profile of a subject, including inferences of underlying health conditions, diseases, and disorders. 

Secondary use scenarios create a unique challenge for privacy, as the data involved are the very same physiological data collected with a user's consent. That is, the physiological data are not being combined with any other source of data. Instead, it is \textit{how} these data are used which extends beyond what users consented to. In a secondary use scenario, a user who has willingly shared their age, sex, and heart rate data to track the intensity of their exercise routine may find that their data could also be used to predict their risk of suffering a heart attack or stroke.

To this end, physiological wearables are likely to create circumstances that introduce unique ethical challenges. On the one hand, informing a user that they have a high risk of suffering a life-threatening injury while it can still be prevented seems to be the right thing to do. On the other, using information a user has provided for purposes beyond what they originally consented to could be considered a violation of their privacy.

\subsection{Auxiliary Use}

The potential for unethical use of physiological signals expands greatly when considering how it may be paired with additional non-physiological data. While restricting access to unnecessary physiological information is more straightforward in direct monitoring applications built for specific purposes such as fitness or mindfulness meditation, for applications and uses with less well-defined boundaries, as well as for general all-day wear/monitoring of physiological signals, the "leaking" of physiological signals may be more difficult to mitigate.

As stated earlier, heart rate, skin conductivity, and certain forms of brain monitoring can be used to detect emotional states. While emotional states may be approximated using other digital signals, such information is much less direct than that which can be offered using physiological wearables. A news organization that promises to offer readers a customized experience that optimizes their well-being by ensuring they are recommended articles that induce a "healthy" variety of emotions in an attempt to circumvent the "dooms-scrolling" phenomenon could instead discover, using readers' physiological signals, the private opinions of their readers with respect to certain content. Thus, readers who opted into such a system to improve their habits may inadvertently find themselves subjected to increasingly sophisticated and surreptitious marketing campaigns for products, whether they be restaurant and food choices or politicians and worldviews.

Naturally, while operating in an information economy, the more information one has (assuming it is of good quality), the greater one's advantage. In this manner, digital organizations that primarily rely upon the collection of user data to turn a profit have every incentive to collect as much information as possible from users and to ensure competitors and other third parties do not get access to this information (at least not for free). Ultimately, the complex interplay between the advantages and liabilities associated with physiological data demands an urgent consideration of the potential consequences that may emerge alongside potential benefits.

\section{Risks and Rewards}
We have identified some of the potential sources of unethical use: the manufacturers of physiological sensors, the app developers that utilize physiological signals directly to drive their software, governments, and policy-makers, and third parties that use data gathered via physiological monitoring for purposes other than directly interfacing with some application. Given the diversity of these independent stakeholders, is it possible to ensure they behave ethically and with careful consideration to the general public, even when unethical and self-interested behavior confers significant material benefits? The initial, instinctive response is "Yes, of course" - after all, we can just enact legislation that forbids unethical behavior, right?

While all have their share of risks and benefits associated with them, these are not necessarily equal in scope or severity. These concerns stem from four general categories: \textit{Awareness and Transparency}, \textit{Biometric Identification}, and \textit{Coerced Utilization}.

\textit{Awareness and Transparency:} Users are unable to make informed decisions regarding their data if they are unaware of \textit{how} their data could be used in the first place. A user who has a typical familiarity with the law also does not necessarily understand the terms of service they agree to upon installing an application or joining a service. Yet, these agreements are regularly used to grant the service provider in question the legal right to more or less do as they please with their users and the data collected from them. 

\textit{Access Control and Surveillance}: While users may be able to hide their information from other users or general members of the public, there are no guarantees that their information is safe from organizations or their benefactors (ranging from interested governments to marketing firms). A commonly expressed user desire is for the ability to delete previously collected data as well as any insights or characteristics that were discovered through these data. However, the ease at which this desire can be exercised may vary significantly depending on the application and what protections a user may have been granted by governing authorities. 
    
\textit{Biometric Identification}: Facial recognition and voice recognition both present privacy concerns for users, particularly in the context of surveillance. However, other forms of biometric identification are possible through certain types of raw physiological signals, such as EEG. Thus any application that relies upon the collection of the appropriate physiological signals may inadvertently expose the identity of a user if it mishandles the collected data. On the other hand, biometric signals may prove much more difficult for scammers to forge, and could therefore prove to be a useful alternative to identity verification for certain services.
    
\textit{Coerced Utilization}: Technologies, where the perceived benefits seem to significantly outweigh their costs, can create dramatic shifts in collective behavior and social norms. Sufficiently transformative technologies can create new expectations where the everyday use of the technology is assumed by default. This can fundamentally change the behaviors of governments, private companies, and social groups in ways where those unwilling or unable to adopt the technology face unique disadvantages and challenges.

\subsection{User Perspectives}
We can begin by focusing on the individuals that may be using these devices. After all, a user base that possesses a healthy level of skepticism and caution towards the purported benefits of physiological computing would place unique pressure on organizations wishing to capitalize on physiological data. Furthermore, users who are fully aware of the value of their physiological data would be less inclined to freely give it away or otherwise grant access privileges to third parties simply out of convenience or impatience. While users may not value all of the risk categories equally, it can be safely assumed that most of the risk categories will be taken into consideration by the user at some point.

Recent work \cite{prange2021investigating} has shown that individuals become more hesitant to use wearables the more they know about what their data can be used for, particularly when it reveals information about them otherwise unrelated to the use case of interest. \textit{Awareness and Transparency} of how data are used and an understanding of how \textit{Access Control and Surveillance} are handled for a wearable and associated applications are likely to be the most important categories users will first consider. Focusing on educating potential users of physiological wearables on the nature and utility of the information they collect may offer stronger incentives for wearable manufacturers and developers to build their products more securely and provide more transparency on how user data is being used.

A typical user willingly exchanges large amounts of their personal information in order to use online services, and this willingness corresponds to high levels of access to digital technology from a young age. What's more, is that this sharing of information occurs \textit{even when users report they would prefer \textbf{not} to}. This phenomenon, also known as the \textit{privacy paradox} \cite{kokolakis2017privacy}, presents unique challenges to creating the proper incentives for web and app developers to behave more ethically regarding how much data they collect from their users. The majority of society may believe that tech companies unfairly profit from users' personal data, but continue to use these services anyway. It may be that these users assume that their data is not that important, or that they will be able to exercise their right to have their data promptly deleted by the provider of whatever service they are using. Sufficiently informed on the prospects of their data being used for \textit{Biometric Identification}, it's likely that users will be much more cautious towards whom they grant access to their physiological data.

From a policy perspective, the most cautious approach is to assume stewards will do whatever is in their own interests provided the users allow them to, either willfully or due to a lack of awareness. While this assumption may only apply to a minority of stewards, creating policies under this assumption as a prophylactic measure is completely sensible given the power of physiological data.

It seems relatively straightforward to assume that a technically knowledgeable public would translate into a cautious and scrupulous user base more likely to demand ethical behavior and stricter data protections from service providers. Yet different attitudes do not necessarily translate to different behaviors: 63\% of active users of social media in the United States believe that social media has an overall negative effect on U.S. society, compared to 69\% of non-users\footnote{https://www.pewresearch.org/fact-tank/2020/10/15/64-of-americans-say-social-media-have-a-mostly-negative-effect-on-the-way-things-are-going-in-the-u-s-today/}. In practice, negative perception of a particular technology is not enough of a deterrent to prevent people from using it. 

Considering these disparities between expressed opinion and demonstrated behavior, is it even possible to nudge the public towards behaviors that facilitate better control and protection of one's personal data? When surveyed, users generally report that they are less willing to share their personal medical information, however, they would still do so if they believed they would receive better recommendations on how to maintain or improve their own health \footnote{https://globaldma.com/consumer-attitudes/}. From these surveys, it is not entirely clear what a typical user considers to be medical information. However, owing to the capabilities of physiological data to predict medical conditions and psychological states, categorizing all data collected from physiological wearables as medical information seems a cautious and logical choice. Educating the public to recognize this and to afford their physiological data special protections commonly ignored for other types of data should be a top priority for fostering a more skeptical and savvy community of users.

Users may consent to data being collected that they would not otherwise consent to if explicitly asked in a straightforward manner. However, dark design patterns push users to give away more data than they ought to, often strictly out of convenience when a user cannot be bothered to menu dive and revoke permissions set as default by the designers of a device or application. An educated public that is informed on how to protect their personal data, particularly in the context of physiological wearables, is all but required for the ethical development and use of these devices.

\textit{Coerced Utilization} is a difficult risk to manage, as the promises of new technologies may blind societies to their potential negative consequences. Of particular concern is the designing or redesigning of social infrastructure that assumes access to a particular technology. Under the auspices of improving the security of critical applications and services, such as online banking, \textit{Biometric Identification} utilizing physiological signals might be used to verify users rather than other methods, such as facial recognition. While biometric identification might offer more security for banking and other sensitive services, inadvertent exposure of user biometric data could result in them being identified across services and platforms where full knowledge of a user's identity serves little benefit to the user.

\subsection{Steward Perspectives}

As the number of users for BCIs and physiological computing devices continues to grow, it can be expected that there will also be significant growth in the number of developers creating applications for such devices and third parties interested in using the collected data for other purposes; there is no guarantee that these developers will all share the same ethical principles regarding what they can and cannot do with their user's data. However, it is not just developers who must exercise extra care in the deployment and operation of physiological wearables. Any entity entrusted with the data collected from these devices or with the legal authority to require or restrict their use (e.g. a government) is subject to the same principles of caution and care.

\textit{Awareness and Transparency}: Owing to the capacities of collected physiological signals to be used in a manner that facilitates the diagnosis of various physical and mental disorders, all physiological data collected from users should be given the same protections and treated with the same care as private medical data. Stewards must therefore provide users with an easily understood summary of how data are used and for what purpose.

The standards of behavior expected of stewards should be similar to the expectations demanded of medical professionals: confidentiality and a do-no-harm code of ethics. Protecting users from inadvertent \textit{Biometric Identification} should be a foundational concern for stewards, particularly among stewards handling EEG data and other signal sources that can expose the identity of a user.

\textit{Access Control and Surveillance}: Stewards that share a user's data with other stewards (whether they are private businesses or government authorities) must state they are doing so in clear language. They must also make it easy for users to make requests to delete their personal physiological data and to promptly fulfill those requests. Such standards of behavior are critical to ensuring users' rights remain properly protected. Constraints that further restrict the capacity of companies and governments to freely trade in physiological data would further discourage the users-as-products business model.

\textit{Coerced Utilization}: Emergent behaviors arise when a technology no longer functions as a niche interest of hobbyists and enthusiasts but is ubiquitous and part of everyday life. The business models that may emerge from always-on physiological wearables being owned and used by people at similar rates as smartphones are difficult to predict, and therefore even more difficult to prepare for. 

For manufacturers and stewards alike, \textit{Coerced Utilization} is of mixed concern, as the consequences of physiological devices becoming an implicit expectation of everyday life also suggest a reliable demand for their products and services. Negative emergent effects and behaviors, such as growth in anti-competitive lobbying to promote or suppress certain wearables or service ecosystems, are likely to increase in prevalence but are not otherwise unique to \textit{Coerced Utilization} circumstances. Nonetheless, the ability of a manufacturer or steward to avoid adopting dual roles as both manufacturer \textit{and} steward is likely to be compromised with ubiquitous physiological wearable usage. 

\section{Discussion and Conclusions}

We have set out here to draw attention to the manifold problems likely to arise with the incautious and widespread adoption of wearable physiological sensors, as well as the unique benefits they provide. The greatest source of risk is not necessarily through any single device or technology itself but from the mass aggregation of physiological data paired with non-physiological data. As sensor technologies continue to evolve and the analytical techniques used to infer additional information from users become more sophisticated, the risks of unscrupulous data stewards misusing these devices and the associated data only increase. It may be possible to mitigate such risks through public education and advances in cyber-security, however, these risks cannot be eliminated entirely due to the evolving nature of physiological monitoring technologies. 

Users of physiological monitoring software are more or less placing their faith in the stewards that utilize this information to use it only in ways that they consent to and otherwise are aware of. In most of the examples we have presented thus far, we assume that stewards are placing the well-being and health of users as their first responsibility and obligation, which therefore guides these stewards' actions.

A primary component that underlies many of the concerns we have outlined here is the role of incentives. Given the scope of possible user insights that could be extracted from physiological data paired with conventional user behavior, it can be expected there will be overwhelming incentives to glean as much information from users via their physiological devices. The fact that most physiological devices operate in an ``always-on" capacity means the opportunities for exploitation are far more extensive than devices and applications that require some form of intentional user interaction. Designing the devices such that there are built-in privacy settings that automatically shut off the device's sensors when a user is using a web browser or social media is one simple but powerful way to reduce opportunities for exploitation.

Legislation such as Europe's GDPR or California's CCPA, at the outset, may offer some safeguards to users with respect to preventing misuse and abuse of their collected physiological data. On the other hand, not all privacy legislation automatically accounts for physiological data. Colorado's Privacy Act was expanded in 2024 via HB24-1058 to specifically account for data of physiological origin and grant suitable protections for citizens. Meanwhile, the European Union's Artificial Intelligence Act includes specific restrictions to applications utilizing physiological and biometric data, applications that use physiological data to influence user behavior without their knowledge or consent are explicitly prohibited by the act.

Such legislation is only effective if stakeholders are compliant and users are knowledgeable of what they are actually agreeing to when they accept a stakeholder's user agreement. Additionally, most legislation only applies to specific regions and economic areas, thus limiting the scope of protection it may offer. This does not mean that drafting legislation is a futile endeavor - but it is crucial to recognize that the capacity for physiological data to be exploited in novel and unexpected ways presents unique and constantly evolving challenges for lawmakers.

Even if global legislation were to be enacted and perfectly enforced, it does little to contend with the consequences of inevitable physiological data breaches. Depending on the signal type, users' identities can be permanently compromised, along with any insights provided by the data concerning their private activities, interests, and tendencies.

Tackling the risks of physiological data breaches without significantly limiting the potential benefits of the data caused by heavy most likely involves a complete reworking of how user data, physiological or otherwise, are collected and stored. Blockchain-based solutions, such as those proposed in \cite{guan2023blockchain}, given the proper modifications, could significantly reduce the capacities for data brokerage firms to exchange and transfer user data in a manner that better preserves user privacy. Requiring all physiological data to be only accessible through smart contracts that prevent all access and analysis outside the specific context or application they were collected for would further reduce the chances of physiological data being accessed illicitly by third parties. It is in the best interests of both private companies as well as individual users for data stewards to adopt standards that significantly reduce both the risk of data leaks and the practical usefulness of such leaked data.

Despite the concerns we have discussed here, physiological computing is an exciting paradigm with enormous potential to have a truly positive impact on society. While large tech companies, thanks to their unique access to even larger amounts of user data, have been able to construct detailed profiles of their users, physiological computing may offer these capabilities directly to the users themselves for the first time. Physiological computing presents the unique opportunity to re-humanize aspects of our digital world and enhance our understanding of the physical one. 

\section*{Acknowledgments}

The work was supported by the Academy of Finland and the Horizon 2020 FET program (CHIST-ERA-20-BCI-001).

\balance

{\small
\bibliographystyle{ieee_fullname}
\bibliography{references}
}

\begin{IEEEbiographynophoto}
{Keith M. Davis III } is a PhD student at the University of Helsinki, Finland. 
His research interests include brain-computer interfaces and machine learning. 
\end{IEEEbiographynophoto}
\begin{IEEEbiographynophoto}
{Tuukka Ruotsalo} is an Associate Professor at the University of Copenhagen, Denmark, and LUT University, Finland. 
His research interests include machine learning and cognitive and physiological computing. 
\end{IEEEbiographynophoto}

\vfill

\end{document}